\documentclass[aps,twocolumn,superscriptaddress,showpacs,showkeys,preprintnumbers,amsmath,amssymb]{revtex4}
\usepackage{txfonts}
\usepackage{dcolumn}
\usepackage{mathrsfs}
\usepackage{bm}
\usepackage{amsmath,amssymb,epsfig,float,graphics}
\usepackage{epstopdf}
\usepackage{amsfonts}
\usepackage{subfigure}
\begin{document}







\title{Dissipation suppression for an Unruh-DeWitt battery with a reflecting boundary}

\author{Xiaobao Liu}
\affiliation{Department of physics and electrical engineering, Liupanshui Normal University, Liupanshui 553004, Guizhou, P. R. China}

\author{Zehua Tian \footnote{Corresponding author, Email:tzh@hznu.edu.cn}}
\affiliation{School of Physics, Hangzhou Normal University, Hangzhou, Zhejiang 311121, P. R. China}

\author{Jiliang Jing\footnote{Corresponding author, Email: jljing@hunn.edu.cn}}
\affiliation{Department of
Physics, Key Laboratory of Low Dimensional Quantum Structures and
Quantum Control of Ministry of Education, and Synergetic Innovation
Center for Quantum Effects and Applications, Hunan Normal
University, Changsha, Hunan 410081, P. R. China}


\begin{abstract}
In the framework of open quantum systems, we study the dynamics of an accelerated quantum battery (QB), modeled as an Unruh-DeWitt detector interacting with a real massless scalar quantum field. The QB is driven by an external classical force acting as a charger. A major challenge in this setup is the environment-induced decoherence, which leads to energy dissipation of the QB. Accelerated motion exacerbates this dissipation, manifesting effects analogous to those experienced by a static QB in a thermal bath in free space, consistent with the Unruh effect. To overcome these challenges, we introduce a reflecting boundary in a space, which modifies the vacuum fluctuations of the field and leads to a position-dependent suppression of dissipation for the Unruh-DeWitt QB. Our analysis reveals that as the QB approaches the boundary, the relevant dissipation is significantly reduced. In particular, when the QB is placed extremely close to the boundary, the dissipation is nearly eliminated, as if the QB were a closed system.
Furthermore, we identify a characteristic length scale associated with the acceleration of QB.
When the distance between the QB and the boundary is much smaller than this scale, the boundary effectively suppresses dissipation, and this suppression effect becomes identical for both an accelerated QB and a static QB in a thermal bath.
Conversely, when the distance is beyond this scale, the suppression effect weakens and manifests a significant difference between these two cases.
Our findings demonstrate the potential of boundary-induced modifications in vacuum fluctuations to effectively suppress dissipation, offering valuable insights for optimizing QB performance. This work paves the way for the development of high-efficiency quantum energy storage systems in the relativistic framework.
\end{abstract}
\pacs{04.62.+v, 03.70.+k, 42.50.Lc, 05.70.-a}
\keywords{Unruh-DeWitt quantum battery, Dissipation, Vacuum fluctuations, The Unruh effect}

\maketitle


\section{Introduction}\label{section1}
 Quantum properties play a crucial role in the behavior of micro- and nano-scale devices.
The thermodynamic aspects of microscopic quantum systems, such as work and heat, have been extensively investigated both theoretically and
experimentally, making significant contributions to the emerging field of quantum thermodynamics~\cite{Esposito2009,Levy2012,Pekola2015,Vinjanampathy2016,
Benenti2017,Bera2019,Carrega2019}.
A key challenge in this field, with profound technological implications, is how to efficiently harness quantum properties to store energy in microscopic systems and release it on demand for power supply. This challenge has naturally motivated research into QBs and the design of optimal energy storage protocols~\cite{Alicki2013,Campaioli2018,Campaioli2023}.
Significant efforts have been devoted to improving QB performance by optimizing key merit figures, such as maximizing work extraction, minimizing charging time, and enhancing average charging efficiency~\cite{Binder2015,Andolina2019,Hovhannisyan2013,Shi2022,Gyhm2024,
Ferraro2018,Campaioli2017,Riccardo2024}. Beyond theoretical investigations, QBs have also been experimentally realized in various systems, including fluorescent organic molecules embedded in microcavities~\cite{Quach2022SA}, IBM quantum chips~\cite{Gemme2022}, transmon qubits~\cite{Hu2022}, star-topology NMR spin systems~\cite{Joshi2022}, and quantum dots~\cite{Wenniger2023}.

In a seminal paper, Alicki and Fannes defined a QB as a d-dimensional system whose internal
Hamiltonian $H_0$ has non-degenerate energy levels $\epsilon_j$~\cite{Alicki2013}
\begin{equation}
H_0=\sum_{j=1}^d \epsilon_j |j\rangle\langle j|,\;\;{\rm with} \;\; \epsilon_{j+1}>\epsilon_j.
\end{equation}
The internal energy of a QB is then given by ${\rm Tr}[H_0 \rho]$, where $\rho$ is the state of the QB.  The charging and discharging (extracted) processes via a cyclic unitary process can be achieved by applying an external time-dependent field $H_1(t)$, where $H_1(t)$ is turned on at time $t=0$ and off at time $t=\tau$. The initial state of the battery is described by a density matrix $\rho(0)$ and the time evolution of QB $\rho(t)$ is obtained from the Liouville-von Neumann equation as
\begin{equation}\label{H}
\frac{d}{dt}\rho(t)=-i[H_0+H_1(t),\rho(t)].
\end{equation}
The energy $W$ extracted by this procedure is measured with respect to the internal Hamiltonian $H_0$,
\begin{equation}
W={\rm Tr}[\rho(0)H_0]-{\rm Tr}[\rho(\tau)H_0].
\end{equation}
where $\rho(\tau)=U(\tau)\rho_0 U^\dagger(\tau)$ is obtained from the solution of Eq.~(\ref{H}) without the consideration of environments, with $U(\tau)$ being the
time-evolution operator expressed in terms of the time-ordering operator ${\cal T}$,
that is, $U(\tau)={\cal T} {\rm exp} \{-i\int_0^\tau ds[H_0+H_1(s)]\}$.
However, every quantum system must be treated as an open system due to inevitable interactions with the surrounding environment, at least subject to vacuum fluctuations on the quantum scale, leading to environment-induced decoherence effects~\cite{Breuer2002,Haroche2006,Weiss2012,Shu2022,Jing1,Jing2,Jing3,Tian2023,Xiao2024}. Such effects reduce stored energy, lower charging efficiency, and accelerate energy dissipation, posing a major challenge to QB longevity~\cite{Pirmoradian2019}. Consequently, this issue has motivated extensive studies on open QBs~\cite{Farina2019,Carrega2020,Zhao2021,Yao2021}.
Although most studies of QBs have been carried out in non-relativistic or inertial frames, relativistic effects are becoming increasingly important for emerging quantum technologies, such as quantum communication in curved spacetime~\cite{Bruschi2014,Jonsson2020}, quantum metrology of relativistic quantum fields~\cite{Ahmadi2014,Tian2015,Liu2018}, and quantum networks operating in accelerated or gravitational environments~\cite{Grochowski2017,Barrett1998}.
In such scenarios, relativistic quantum field effects, such as vacuum fluctuations and the Unruh effect, can substantially impact the charging dynamics, the energy storage, and dissipation of QBs. Therefore, studying QBs in a relativistic framework provides a valuable platform for exploring the interplay between quantum thermodynamics, general relativity, and quantum field theory.
Recent research has explored the impact of spacetime properties and field boundary conditions on QB dynamics. For example, studies have examined an open QB in BTZ spacetime, shedding insights into the interplay between gravitational effects and QB charging efficiency~\cite{Tian2024}.
Other studies have shown that accelerated motion will suppress energy extraction in a QB, which is interacted with a scalar field through linear or quadratic coupling~\cite{Hao2023,Hao2025,Mukherjee2024}. Intriguingly, certain trajectories, characterized by a combination of linear acceleration and components of the four-velocity \cite{LiuPRD2021, Abdolrahimi2014}, can mitigate energy dissipation. Additional schemes, including feedback control \cite{Yao2022}, dark states \cite{Quach2020}, Floquet engineering \cite{Bai2020}, and environment engineering \cite{Tabesh2020,Xu2021,Ghosh2021,Carrasco2022,Mayo2022,Song2022}, have also been proposed to combat decoherence effects. Despite these efforts, developing effective protection methods to suppress energy dissipation in different types of QB, such as Unruh-DeWitt QB, remains an open issue.

As is well known, the Unruh-DeWitt detector is a widely used and analytically tractable model in relativistic quantum field theory~\cite{DeWitt1979,Unruh1984}. Unlike spin chains or cavity-QED systems~\cite{Campaioli2023}, the Unruh-DeWitt detector model provides a simple yet powerful framework that not only captures essential features of the quantum field, but also admits exact analytical solutions~\cite{Birrell1984,Takagi1986}. As a result, it has been extensively used in studies of the detection of spacetime properties~\cite{Crispino2008,Louko2008,Conroy2022}, entanglement harvesting~\cite{Pozas2015}, and response functions in open quantum dynamics~\cite{Sriramkumar1999,Lin2006}, just to name a few.
Therefore, in this paper, we consider an accelerated QB modeled as an Unruh-DeWitt detector interacting with a massless scalar field. However, the interaction between the moving QB and the quantum field will lead to the decoherence and dissipation of the battery~\cite{Doukas2009}.
To suppress this energy dissipation, we introduce a reflecting boundary. This is a well-established approach in quantum field theory.
Reflecting boundaries have been shown to significantly influence various quantum processes, including the suppression or enhancement of spontaneous excitation rates~\cite{Yu2005,Yu2007}, Lamb shifts~\cite{Yu2010}, the radiation of uniformly accelerated atoms~\cite{Arias2016}, and the preservation of quantum coherence and entanglement~\cite{1Liu2016,2Liu2016,Hu2021}. In our relativistic QB model, since the boundary modifies the vacuum fluctuations of the scalar field, it alters the decay rates of the QB, thereby effectively suppressing energy dissipation.
Our analysis demonstrates that the energy dissipation rates of the QB strongly depend on its distance $z$ from the reflecting boundary. By tuning $z$, the energy dissipation of the QB can be effectively controlled. As $z$ decreases, dissipation is significantly suppressed compared with the free-space case. In particular, when the QB is placed extremely close to the boundary, the energy dissipation caused by environment-induced decoherence and the Unruh effect is nearly eliminated, as if the QB were an isolated system.
Moreover, we introduce a characteristic length scale, $z_a$, associated with the acceleration of QB. When $z\ll z_a$,
the suppression of dissipation remains robust, and the boundary-induced suppression effect for the accelerated QB is indistinguishable from that for a static QB in a thermal bath.  In contrast, when $z \gg z_a$, the suppression effect diminishes, and the boundary-induced suppression effect for the accelerated QB differs significantly from that for a static QB in a thermal bath. Consequently, this boundary-induced shielding represents a promising mechanism for optimizing QB performance.

The structure of our paper is as follows. In Sec. II, we introduce the QB coupled to a massless scalar field and define the maximum of extractable energy, i.e., the ergotropy, in the framework of open quantum systems. In Sec. III, we investigate the impacts of vacuum fluctuations and the Unruh effect on the ergotropy both in free space and in the presence of a reflecting boundary. We then compare the ergotropy of the accelerated QB with that of a static QB in a thermal bath at temperature
$T=a/2\pi$. Finally, in Sec. IV, we present our conclusions.
Throughout the paper, we use natural units $c=\hbar=1$. Relevant constants are restored when needed for the sake of clarity.

\section{Open quantum battery and ergotropy}\label{sectionII}
In this section, we are going to present the model of an open QB coupled to a massless scalar field and the relevant physical quantity to characterize how efficiently energy can be extracted from the QB.

\subsection{Open quantum battery}\label{Dynamics}
We sketch a proposal for a QB modeled as a two-level system, usually known as the Unruh-DeWitt detector.  This system comprises an excited state $|e\rangle$ and a ground state $|g\rangle$, with a transition frequency $\omega_0$ between the two levels. The Hamiltonian of this QB is expressed as $H_b=\frac{\omega_0}{2} \sigma_z$.
To implement the charging process, the classical coherent field can be utilized to drive the QB; thus, at time $\tau=0^+$, the quantum battery can be
charged by applying a local external driving field in the $x$ direction with strength $\omega$~\cite{Ghosh2020,Huang2023}, as
$H_c=-\frac{\omega}{2} \sigma_x$. In the framework of QB, the parameter $\tau$ signifies the proper charging time.
Additionally, assume that the QB is coupled to a fluctuating massless scalar field $\Phi(x(\tau))$ in vacuum, which may result in the decoherence of the QB. The Hamiltonian of the combined system (battery + external field) can be written as follows:
\begin{equation}\label{Hamiltonian0}
H=H_{b}+H_{c}+H_{f}+H_I,
\end{equation}
where $H_{f}$ is the Hamiltonian of the scalar field and its explicit expression is not required here.
The QB-field coupling is assumed to be linear along the transverse $(x)$ direction as
\begin{equation}
H_I=\mu\sigma_x\Phi(x(\tau)),
\end{equation}
where $\mu$ is the coupling constant that we assume to be small.
In order to investigate the dynamical evolution of the QB, we consider a unitary rotation in the spin space, represented as $R(\Theta)=e^{-\textrm{i}\frac{\Theta}{2}\sigma_y}$. Here, the angle $\Theta$ is carefully selected to ensure that the combined QB-charger Hamiltonian $H_{b}+H_{c}$ is projected only along the $z$-axis.
Under unitary rotation, we can obtain the following:
\begin{align}\label{Hamiltonian0}
&\tilde{H}_b=RH_bR^\dagger=\frac{\omega_0}{2} (\cos\Theta\sigma_z+\sin\Theta\sigma_x),\nonumber\\
&\tilde{H}_c=RH_cR^\dagger=\frac{\omega}{2} (\sin\Theta\sigma_z-\cos\Theta\sigma_x),\\
&\tilde{H}_{bc}=\tilde{H}_b+\tilde{H}_c=\frac{\Omega}{2}\sigma_z,\nonumber\\
&\tilde{H}_I=RH_IR^\dagger=\mu[\cos\Theta\sigma_x-\sin\Theta\sigma_z]\Phi(x(\tau)),\nonumber
\end{align}
where $\cos\Theta=\omega_0/\Omega$, $\sin\Theta=\omega/\Omega$ and $\Omega^2=\omega_0^2+\omega^2$.

Initially, the state of QB and the quantum field can be written as the density matrix $\rho(0)\otimes|0\rangle\langle0|$. Here, $\rho(0)=\frac{1}{2}[\mathbf{I}+\vec{r}(0)\cdot \boldsymbol{\sigma}]$ is the reduced density matrix of QB with $\boldsymbol{\sigma}=(\sigma_1,\sigma_2,\sigma_3)$, and $|0\rangle$ is the vacuum of the scalar field defined by $a_k|0\rangle=0$ for all $k$.
Therefore, the rotated reduced initial matrix of the QB is $\tilde{\rho}(0)=R(\Theta)\rho(0)R(\Theta)^\dagger=\frac{1}{2}[\mathbf{I}+\vec{\tilde{r}}(0)\cdot \boldsymbol{\sigma}]$, satisfying
\begin{align}
&\tilde{r}_1(0)=r_1(0)\cos\Theta+r_3(0)\sin\Theta,\nonumber\\
&\tilde{r}_2(0)=r_2(0),\\
&\tilde{r}_3(0)=r_3(0)\cos\Theta-r_1(0)\sin\Theta.\nonumber
\end{align}
For the total system, its equation of motion in the interaction picture is
\begin{equation}
\frac{\partial\tilde{\rho}_{\textrm{tot}}(\tau)}{\partial\tau}=-\textrm{i}[\tilde{H}_I(\tau),\ \tilde{\rho}_{\textrm{tot}}(\tau)].
\end{equation}
Since our focus is on the dynamics of the QB, we trace over the field's degrees of freedom.
Given the weak interaction between the QB and the quantum field and by applying the Born approximation and the Markov approximation~\cite{Breuer2002}, the master equation describing the dissipative dynamics of the QB can be expressed in the Kossakowski-Lindblad form~\cite{Manzano2020,Benatti2003}
\begin{equation}\label{Lindblad equation}
\frac{\partial\tilde{\rho}(\tau)}{\partial\tau}=-\textrm{i}[\tilde{H}_{{\rm eff}},\tilde{\rho}(\tau)]+{\cal L}[\tilde{\rho}(\tau)],
\end{equation}
where
\begin{equation}\label{L}
{\cal L}[\tilde{\rho}]=\frac{1}{2}\sum^3_{i,j=1}a_{ij}\big[2\sigma_j\tilde{\rho} \sigma_i-\sigma_i \sigma_j\tilde{\rho}-\tilde{\rho}  \sigma_i \sigma_j\big]
\end{equation}
is the dissipator. From the master equation~(\ref{Lindblad equation}), it is clear that the environment leads to decoherence and dissipation described by the dissipator ${\cal L}[\tilde{\rho}(\tau)]$, such that the evolution of the quantum system is non-unitary
and also gives rise to a modification of the unitary evolution term which incarnates in the Hamiltonian $H_{{\rm eff}}$.  The Kossakowski matrix $a_{ij}$ in the dissipator (\ref{L}) can be written as
\begin{equation}
a_{ij}=A \delta_{ij}-\textrm{i} B\epsilon_{ijk} \delta_{k3}
+C \delta_{i3} \delta_{j3},
\end{equation}
with
\begin{align}
&A=\frac{1}{2}\big[{\cal G}(\Omega)+{\cal G}(-\Omega)\big],\;\;\nonumber\\
&B=\frac{1}{2}\big[{\cal G}(\Omega)-{\cal G}(-\Omega)\big],\\
&C=\frac{1}{2}\big[2{\cal G}(0)-{\cal G}(\Omega)-{\cal G}(-\Omega)\big].\nonumber
\end{align}
We introduce the two-point Wightman function for the scalar field as
$G^+(x(\tau),x(\tau'))=\langle0|\Phi(x(\tau))\Phi(x(\tau'))|0\rangle$,
and its Fourier and Hilbert transforms of the field Wightman
function read, respectively, as follows:
\begin{align}\label{transforms0}
&{\cal G}(\lambda)=\mu^2\cos^2\Theta\int^{+\infty}_{-\infty} d\Delta\tau e^{ \textrm{i} \lambda \Delta\tau}G^+(x(\tau),x(\tau')),\nonumber\\
&{\cal G}(0)=\mu^2\sin^2\Theta\int^{+\infty}_{-\infty} d\Delta\tau G^+(x(\tau),x(\tau')),\\
&{\cal K}(\lambda)=\frac{P}{\pi \textrm{i}}\int^{+\infty}_{-\infty} d\omega' \frac{{\cal G}(\omega')}{\omega'-\lambda},\nonumber
\end{align}
where $P$ represents the principal value and $\Delta\tau=\tau-\tau'$.
By absorbing the Lamb shift term, the effective Hamiltonian $H_{{\rm eff}}$ can be written as
\begin{equation}\label{Effective H}
\tilde{H}_{{\rm eff}}=\frac{1}{2}\Omega'\sigma_z
=\frac{1}{2}\bigg\{\Omega+\textrm{i}\big[{\cal K}(-\Omega)-{\cal K}(\Omega)\big]\bigg\}\sigma_z,
\end{equation}
where $\Omega'$ denotes the effective energy level-spacing of the QB with a correction term $\textrm{i}\big[{\cal K}(-\Omega)-{\cal K}(\Omega)\big]$ being the Lamb shift. Usually, the Lamb shift can be neglected because it is much less than $\Omega$, that is, $\Omega'\approx\Omega$.

For a two-level QB, the state of the system can be expressed in the Bloch-sphere representation as follows:
\begin{equation}\label{initial state}
\tilde{\rho}=\frac{1}{2}\bigg[\mathbf{I}+\vec{\tilde{r}}\cdot\boldsymbol{\sigma}\bigg],
\end{equation}
where $\mathbf{I}$ is the $2\times2$ unit matrix, $\vec{\tilde{r}}=(\tilde{r}_1,\tilde{r}_2,\tilde{r}_3)$ represents the Bloch vector. Inserting the Eq.~(\ref{initial state}) into Eq.~(\ref{Lindblad equation}), the master equation compatible with complete positivity of time-evolution is equivalent to the three coupled differential equations satisfying
\begin{align}\label{evolution equation}
&\frac{\partial \tilde{r}_1}{\partial\tau}=-2(2A+C)\tilde{r}_1-\Omega \tilde{r}_2,\nonumber\\
&\frac{\partial \tilde{r}_2}{\partial\tau}=\Omega\tilde{r}_1-2(2A+C)\tilde{r}_2,\\
&\frac{\partial \tilde{r}_3}{\partial\tau}=-4A\tilde{r}_3-4B.\nonumber
\end{align}
As shown in Ref.~\cite{Alicki1987}, the evolution equation in Eq.~(\ref{evolution equation}) can be conveniently rewritten as a Schr\"{o}dinger-like equation as
\begin{equation}\label{Bloch vector}
\frac{\partial}{\partial\tau}\vec{\tilde{r}} (\tau)=-2{\cal H}\cdot \vec{\tilde{r}} (\tau) + \vec{\chi},
\end{equation}
where $\vec{\chi}^T=(0,0,-4B)$ denotes the inhomogeneous vector, which comes from the
imaginary part of the Kossakowski matrix $a_{ij}$, and the decaying matrix ${\cal H}$ takes the form of
\begin{equation}\label{Hmatrix}
{\cal H}=
\left(
\begin{array}{ccc}
2A+C& \Omega/2& 0\\
-\Omega/2& 2A+C& 0\\
0& 0& 2A
\end{array}
\right).
\end{equation}
which is the $3\times3$ matrix that includes contributions from both $H_{{\rm eff}}$ and the real part of $a_{ij}$.
Let us assume the initial reduced density matrix of the QB which is expressed as $\tilde{\rho}(0)$,
and submitting $\tilde{\rho}(0)$ into eq.~({\ref{evolution equation}}),
then the time-dependent state
parameters $\tilde{\rho}(\tau)$ are found to be
\begin{align}\label{PM2-0}
\tilde{r}_1(\tau)&=\tilde{r}_1(0)\cos(\Omega\tau) e^{-2(2A+C)\tau}\nonumber\\
&-\tilde{r}_2(0)\sin(\Omega\tau) e^{-2(2A+C)\tau},\nonumber\\
\tilde{r}_2(\tau)&=\tilde{r}_1(0)\sin(\Omega\tau) e^{-2(2A+C)\tau}\\
&-\tilde{r}_2(0)\cos(\Omega\tau) e^{-2(2A+C)\tau},\nonumber\\
\tilde{r}_3(\tau)&=\tilde{r}_3(0)e^{-4A\tau}+\frac{B}{A}(e^{-4A\tau}-1).\nonumber
\end{align}
Rotating $\tilde{\rho}$ in eq.~(\ref{initial state}) with state parameters in eq.~(\ref{PM2-0}),
we can get the evolution state of the QB, namely, $\rho(\tau)=R(\Theta)^\dagger \tilde{\rho}(\tau) R(\Theta)$.
Initially, if the QB state was prepared in the ground state $|g\rangle$, that is, $r_1(0)=r_2(0)=0, r_3(0)=-1$, the evolution of QB in eq.~(\ref{PM2-0}) can then be obtained as follows:
\begin{align}\label{PM2}
&\tilde{r}_1(\tau)=-\sin\Theta\cos(\Omega\tau) e^{-2(2A+C)\tau},\nonumber\\
&\tilde{r}_2(\tau)=-\sin\Theta\sin(\Omega\tau) e^{-2(2A+C)\tau},\\
&\tilde{r}_3(\tau)=-\cos\Theta e^{-4A\tau}+\frac{B}{A}(e^{-4A\tau}-1).\nonumber
\end{align}

\subsection{Ergotropy}\label{ergotropy}
Now, we are interested in characterizing how efficiently energy can be extracted from a QB. The energy of the QB can be obtained as
\begin{equation}\label{energy}
E(\tau)=\mathrm{Tr}[H_b \rho(\tau)]=\mathrm{Tr}[\tilde{H}_b \tilde{\rho}(\tau)].
\end{equation}
According to the second law of thermodynamics, not all energy $E(\tau)$ can be converted into work~\cite{Monsel2020}.
The maximum of the extractable energy called ergotropy is defined as~\cite{Allahverdyan2004,Francica2020,Cakmak2020}
\begin{equation}\label{ergotropy0}
\mathcal{W}(\tau)=E(\tau)-\mathrm{min}_{\{U\}}\mathrm{Tr}[U\rho(\tau)U^\dag  H_b].
\end{equation}
The minimal unitary transformation $U_\sigma$ satisfies the condition that states $U_\sigma\rho(\tau)U_\sigma^\dag=\sum_j \varrho_j |\varepsilon_j\rangle\langle\varepsilon_j|$
are passive. Here, $\varrho_j$ are the eigenvalues of $\rho(\tau)$ ordered in a descending sort, and $|\varepsilon_j\rangle$ are the eigenstates of $H_b$ with the corresponding
eigenvalues $\varepsilon_j$ ordered in an ascending sort.
Note that we can obtain $\mathrm{Tr}[U_\sigma\rho(\tau)U_\sigma^\dag H_b]=\mathrm{Tr}[\tilde{U}_\sigma\tilde{\rho}(\tau)\tilde{U}_\sigma^\dag \tilde{H}_b]$ with
$\tilde{U}_\sigma\tilde{\rho}(\tau)\tilde{U}_\sigma^\dag=\sum_j \tilde{\varrho}_j |\epsilon_j\rangle\langle\epsilon_j|$.
Here, $|\epsilon_j\rangle$ is the state of the energy level of $\tilde{H}_b$, and $\{\tilde{\varrho}_j,\;j=1,2\}$ are the eigenvalues of $\tilde{\rho}(\tau)$ calculated as $\tilde{\varrho}_{1,2}=\frac{1}{2}[1\pm \tilde{r}]$,  with $\tilde{r}=|\vec{\tilde{r}}|$ being the norm of the Bloch vector.
Then, the maximum extractable energy of the QB can be achieved by the optimal unitary operation
$\tilde{U}_\sigma=\Sigma_{j=1,2}|\epsilon_j\rangle\langle \tilde{\rho}_j |$, where $|\tilde{\rho}_j\rangle$ is the eigenvector of the
density matrix.
Now, we work on the basis
$\{|\epsilon_1\rangle=|g\rangle,|\epsilon_2\rangle=|e\rangle\}$, and the eigenvectors corresponding to eq.~(\ref{initial state}) can be derived as
\begin{align}\label{eigenvector}
&|\tilde{\rho}_1\rangle=\sqrt{\frac{\tilde{r}+\tilde{r}_3}{2\tilde{r}}}|g\rangle
+\frac{\tilde{r}_1+\textrm{i}\tilde{r}_2}{\sqrt{2\tilde{r}(\tilde{r}+\tilde{r}_3)}}|e\rangle,\\
&|\tilde{\rho}_2\rangle=\frac{\tilde{r}_1-\textrm{i}\tilde{r}_2}{\sqrt{2\tilde{r}(\tilde{r}+\tilde{r}_3)}}
|g\rangle-\sqrt{\frac{\tilde{r}+\tilde{r}_3}{2\tilde{r}}}|e\rangle.\nonumber
\end{align}
Therefore, the optimal unitary transformation can be expressed as:
\begin{equation}\label{transformation}
\tilde{U}_\sigma=
\left(
\begin{array}{cc}
\sqrt{\frac{\tilde{r}+\tilde{r}_3}{2\tilde{r}}}
&\frac{\tilde{r}_1-i\tilde{r}_2}{\sqrt{2\tilde{r}(\tilde{r}+\tilde{r}_3)}} \\
\frac{\tilde{r}_1+i\tilde{r}_2}{\sqrt{2\tilde{r}(\tilde{r}+\tilde{r}_3)}}
&-\sqrt{\frac{\tilde{r}+\tilde{r}_3}{2\tilde{r}}}\\
\end{array}
\right).
\end{equation}

Plugging eqs.~(\ref{PM2}) and  (\ref{transformation}) into eq.~(\ref{ergotropy0}),
the analytical expression of the ergotropy $\mathcal{W}$ for the QB can be expressed as follows~\cite{Farina2019}
\begin{align}\label{scaledergotrogy0}
\mathcal{W}(\tau)&=\frac{\omega_0}{2}\bigg[\sqrt{[-\frac{\omega_0}{\Omega} e^{-J\tau}+\frac{B}{A}(e^{-J\tau}-1)]^2
+\frac{\omega^2}{\Omega^2} e^{-2S\tau}}\nonumber\\
&-\frac{\omega_0^2}{\Omega^2} e^{-J\tau}+\frac{\omega_0}{\Omega}\frac{B}{A}(e^{-J\tau}-1)\nonumber\\
&-\frac{\omega^2}{\Omega^2} e^{-S\tau}\cos(\Omega\tau)\bigg].
\end{align}

Note that in eq.~(\ref{scaledergotrogy0}), the factor $J=4A$ represents the incoherent relaxation rate, which determines how quickly energy dissipation occurs due to the interaction between the QB and the external environment.
In addition, the factor $S=2(2A+C)$ represents the dephasing rate, which governs the loss of quantum coherence in the QB.
Without considering environment, that is, taking QB as a closed system, whose evolution is governed by the Hamiltonian $\tilde{H}_{bc}$, one can find that the Bloch vector of the state with time $\tau$ is $\vec{\tilde{r}}(\tau)=(-\sin\Theta \cos(\Omega\tau),-\sin\Theta\sin(\Omega\tau),-\cos\Theta)$. In this respect,
the ergotropy for a closed QB is found to be
\begin{equation}\label{ergotrogy-closed}
\mathcal{W}(\tau)=\frac{\omega_0}{2}\frac{\omega^2}{\Omega^2}[1-\cos(\Omega\tau)],
\end{equation}
which is a periodic oscillation with time.


\section{Influences of vacuum fluctuations and the Unruh effect on ergotropy} \label{sectionIII}
In this section, we examine how vacuum fluctuations and the Unruh effect influence the ergotropy of an Unruh-DeWitt QB, both in free space and in the presence of a reflecting boundary. We then compare these results with those of a static QB immersed in a thermal bath.

\subsection{Ergotropy of Unruh-DeWitt QB in free space}\label{emptyspace}
We first evaluate the ergotropy of an Unruh-DeWitt QB interacting with a
vacuum massless scalar field in free space. In this scenario, the QB undergoes a uniform acceleration along the $x$-axis with a proper acceleration $a$. The associated trajectory is given by
\begin{equation}\label{trajectory0}
t(\tau)=\frac{1}{a}\sinh a\tau,\;\;\;x(\tau)=\frac{1}{a}\cosh a\tau,\;\;\; z(\tau) =y(\tau)= 0.
\end{equation}
The Wightman function for a massless scalar field in a Minkowski vacuum is given by~\cite{Birrell1984}
\begin{align}\label{correlated function0}
&G_0^+(x(\tau),x(\tau'))
\nonumber\\
&=-\frac{1}{4\pi^2}
\frac{1}{(t-t'-i\epsilon)^2-(x-x')^2-(y-y')^2-(z-z')^2},\nonumber\\
\end{align}
where $\epsilon\rightarrow 0^+$, and the subscript $0$ denotes the Wightman function in the free space.
Substituting the trajectory of QB eq.~(\ref{trajectory0}) into Eq.~(\ref{correlated function0}), we obtain
\begin{equation}\label{correlated function1}
G_0^+(x(\tau),x(\tau'))=-\frac{1}{16\pi^2}\bigg[\frac{a^2}{\sinh^2(\frac{a\Delta\tau}{2}-i\epsilon)}
\bigg],
\end{equation}
where $\Delta\tau=\tau-\tau'$.
According to Eqs.~(\ref{transforms0}) and (\ref{correlated function1}), the Fourier transforms can be evaluated as follows:
\begin{equation}\label{Fourier transforms}
{\cal G}(\lambda)=\frac{\mu^2\omega_0^2}{2\pi\Omega^2}\frac{\lambda}{1-e^{-2\pi\lambda/a}},\;\;\;\;
{\cal G}(0)=\frac{\mu^2\omega^2a }{4\pi^2\Omega^2}.
\end{equation}
The coefficients of the Kossakowski matrix $a_{ij}$ are now given by
\begin{align}\label{ABa1}
&A=\frac{\Gamma\omega_0^2}{2\Omega^2}\coth\bigg(\frac{\pi \Omega}{a}\bigg),
\;\;\;\;B=\frac{\Gamma\omega_0^2}{2\Omega^2},\\
&C=\frac{\Gamma}{2}\bigg[\frac{\omega^2}{\Omega^2}\frac{a}{\pi\Omega}-\coth\bigg(\frac{\pi \Omega}{a}\bigg)\frac{\omega_0^2}{\Omega^2}\bigg],\nonumber
\end{align}
where $\Gamma=\frac{\mu^2\Omega}{2\pi}$ is the spontaneous emission rate of QB in free space.
Substituting Eq.~(\ref{ABa1}) into (\ref{PM2}), one finds that when evolving long enough time, the asymptotic density matrix of QB $\rho_\infty$ is
\begin{equation}\label{eventually sate}
\rho_\infty=\frac{e^{-\beta H_b}}{\mathrm{Tr}[e^{-\beta H_b}]},
\end{equation}
which is a thermal state with a temperature $T=1/\beta=a/2\pi$. Therefore, when a two-level uniformly accelerated particle detector is coupled to a massless scalar field, this detector feels as if it were immersed in a thermal bath with temperature $T=a/2\pi$. This is a manifestation of the well-known Unruh effect~\cite{Unruh1976,Fulling1973,Davies1975,Birrell1984}.

In the following discussion, we will work with dimensionless quantities by rescaling time $\tau$ and acceleration $a$,
\begin{equation}
\tilde{\tau}=\Gamma \tau,\;\;\;\;\;\;\tilde{a}=\frac{a}{\Omega}.
\end{equation}
For simplicity, hereafter we refer to $\tilde{\tau}$ and $\tilde{a}$ as $\tau$ and $a$, respectively.
As a result,
we obtain the detailed formula for the ergotropy of the QB in free space as follows:
\begin{align}\label{ergotropy1}
&\mathcal{W}_0(\tau)=\frac{\omega_0}{2}\nonumber\\
&\times\bigg[\sqrt{[-\frac{\omega_0}{\Omega} e^{-J_0\tau}+\coth^{-1}\bigg(\frac{\pi}{a}\bigg)(e^{-J_0\tau}-1)]^2
+\frac{\omega^2}{\Omega^2} e^{-2S_0\tau}}\nonumber\\
&-\frac{\omega_0^2}{\Omega^2} e^{-J_0\tau}+\frac{\omega_0}{\Omega}\coth^{-1}\bigg(\frac{\pi}{a}\bigg)(e^{-J_0\tau}-1)\nonumber\\
&-\frac{\omega^2}{\Omega^2} e^{-S_0\tau}\cos(\Omega\tau)\bigg],
\end{align}
where the incoherent relaxation and dephasing rates, respectively, are
\begin{equation}\label{rate0}
J_0=\frac{2\omega_0^2}{\Omega^2}\coth\bigg(\frac{\pi}{a}\bigg),
\;\;\;\;\;\;S_0=\frac{J_0}{2}+\frac{\omega^2a}{\Omega^2\pi}.
\end{equation}
It is easy to find that for small accelerations (low Unruh temperature), the energy dissipation is weak, while for high accelerations it increases significantly, leading to faster decoherence. The additional term $\omega^2a/(\Omega^2\pi)$ shows that as acceleration increases, the loss of phase coherence is also enhanced, reducing the efficiency of quantum energy storage. This occurs because the accelerated QB experiences a Minkowski vacuum of a quantum field as an effective thermal bath, which leads to noise that disturbs energy extraction.

\begin{figure}[H]
\centering
{\includegraphics[height=1.87in,width=3.09in]{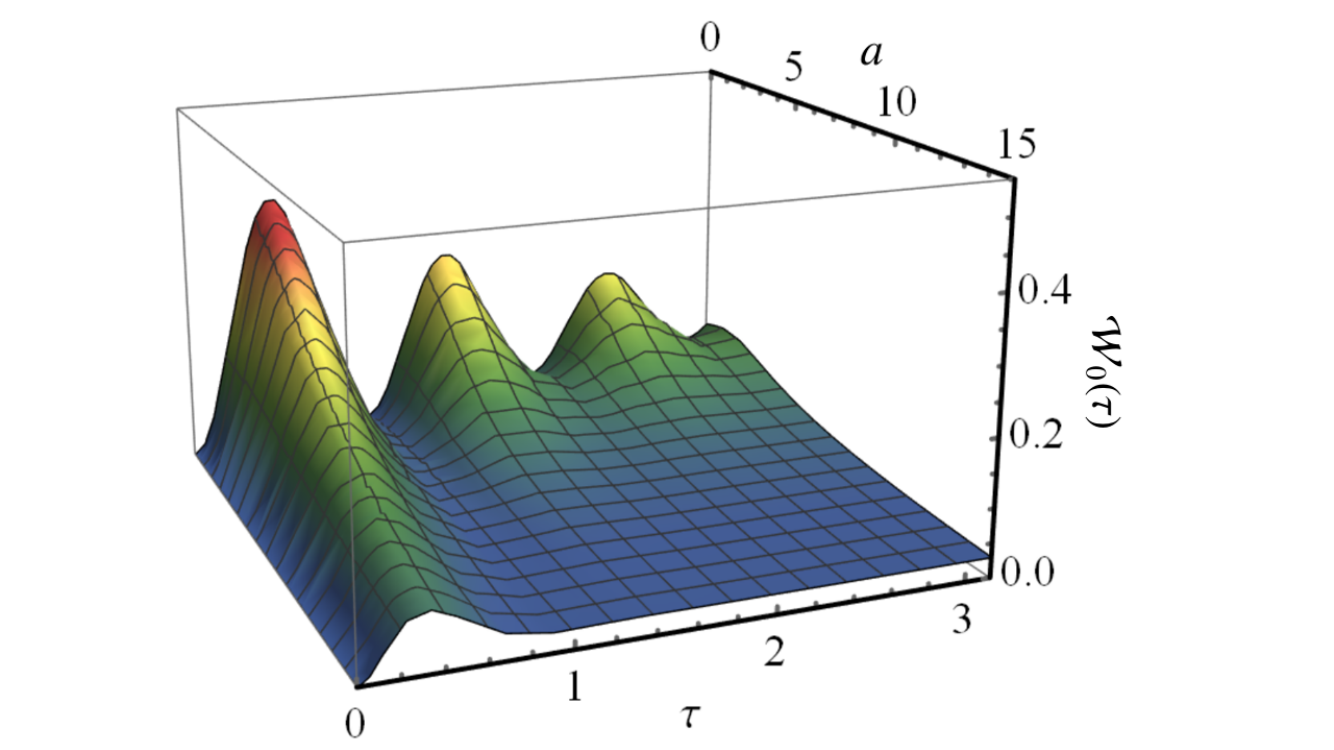}}
\caption{Ergotropy $\mathcal{W}_0(\tau)$ (per unit $\omega_0$) versus time $\tau$ and acceleration $a$ in free space. Here, we fixed $\omega=1$ in the unit $\omega_0$.
}\label{figure0}
\end{figure}

We now analyze the evolution of the ergotropy (per unit $\omega_0$) in free space as a function of time $\tau$ and acceleration $a$, as plotted in Fig.~\ref{figure0}.
During a short charging time, the cyclical unitary operation can contribute to the increase in energy extraction, since the external coherent field transfers the energy from the charger to the battery. However, with increasing acceleration, the incoherent relaxation and dephasing rates in Eq.~(\ref{rate0}) are enhanced, which result in that the ergotropy gradually decreases. This implies that the Unruh effect will expedite the energy dissipation from the dynamics, owing to the interaction between the accelerated QB and the environment.
In the absence of battery-environment coupling, the ergotropy evolves periodically in time, as given by Eq.~(\ref{ergotrogy-closed}), reflecting purely unitary dynamics of a closed QB. However, when the QB interacts with the quantum field, the environment induces both relaxation and dephasing, leading to dissipative evolution. As a result, the ergotropy over time presents a damped oscillation, which arises from the decay rates determined by the Kossakowski matrix. Similar results have been reported in other two-level QB models~\cite{Hao2025,Carrega2020,Tian2024,Hao2023,Mukherjee2024}.
Ultimately, in the infinite time limit, that is, $\tau\rightarrow\infty$, the ergotropy approaches the steady nonzero value:
\begin{equation}\label{ergotropy1-1}
\mathcal{W}(\infty)=
\frac{\omega_0}{2}\bigg(1-\frac{\omega_0}{\Omega}\bigg)\coth^{-1}\bigg(\frac{\pi}{a}\bigg),
\end{equation}
which decreases monotonically with increasing acceleration $a$. This implies that the Unruh effect enhances energy dissipation, reducing the QB's energy storage capacity. The presence of a thermal-like bath as a result of acceleration fundamentally alters the dissipative dynamics, leading to the challenge of energy storage in relativistic QBs.

\subsection{Ergotropy of Unruh-DeWitt QB with a reflecting boundary}\label{boundaryspace1}
\begin{figure}[H]
\centering
{\includegraphics[height=1.85in,width=3.15in]{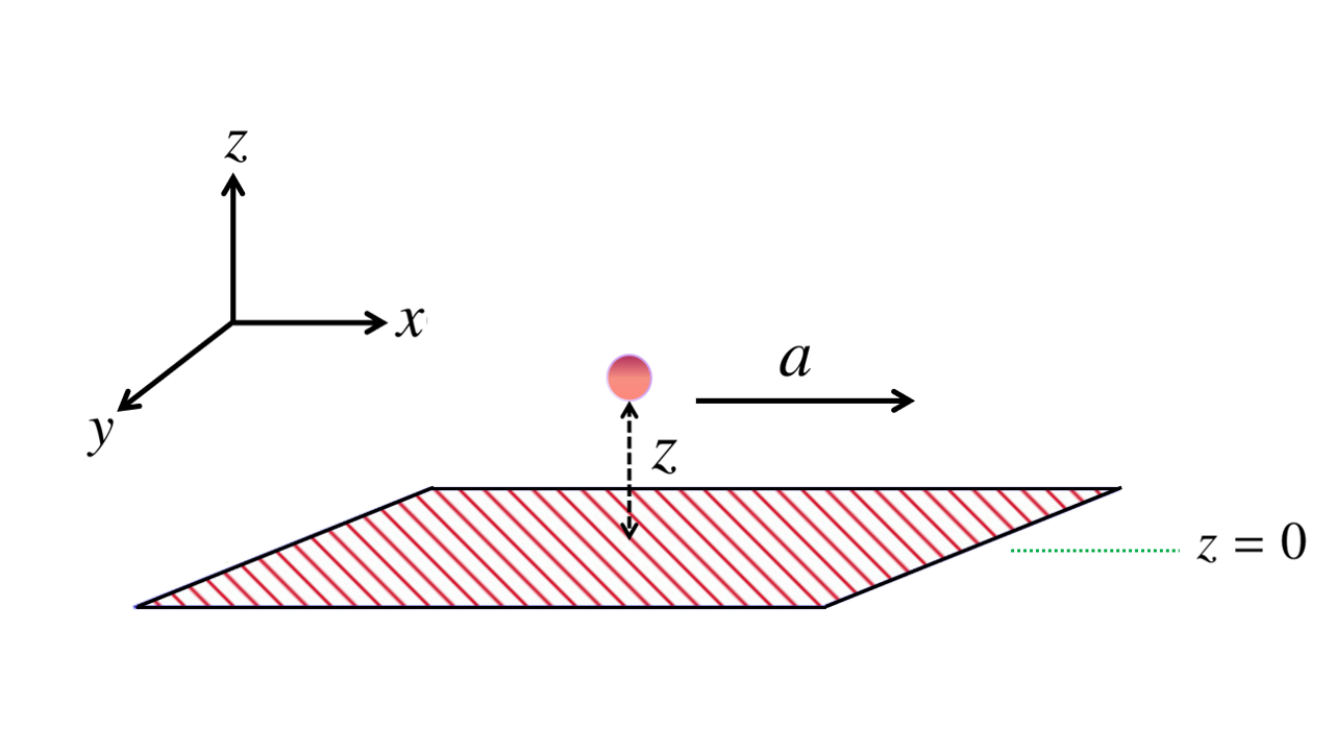}}
\caption{ Uniformly accelerated QB is at a distance $z$ from the infinite reflecting boundary.
}\label{boundary}
\end{figure}

Since vacuum fluctuation will be modified if we set a boundary in the vacuum, it is natural to investigate how a reflecting boundary affects the ergotropy of an Unruh-DeWitt QB. In the following, we consider a QB that interacts with a massless scalar field while undergoing uniform acceleration parallel to the boundary. As depicted in Fig.~\ref{boundary}, the QB maintains a constant distance $z$ from the boundary and accelerates uniformly along the $x$-direction.
In our model, the boundary at $z=0$ is modeled as an ideal infinite plane that imposes Dirichlet boundary conditions on the scalar field~\cite{Birrell1984,Casimir1948}.
A finite-size boundary would introduce additional physical effects such as edge diffraction, mode leakage, and spatial non-uniformity
\cite{Barton2001,Scheel2008}, which are not considered in this work.
Note that $z=0$ denotes the location of the reflecting boundary, not the QB position. In experimental implementations, reflecting boundaries can be engineered using conducting plates, superconducting interfaces, or Casimir-type cavities~\cite{Klimchitskaya2006,Munday2009,Klimchitskaya2009}.

\begin{figure}[H]
\centering
{\includegraphics[height=2.05in,width=3.35in]{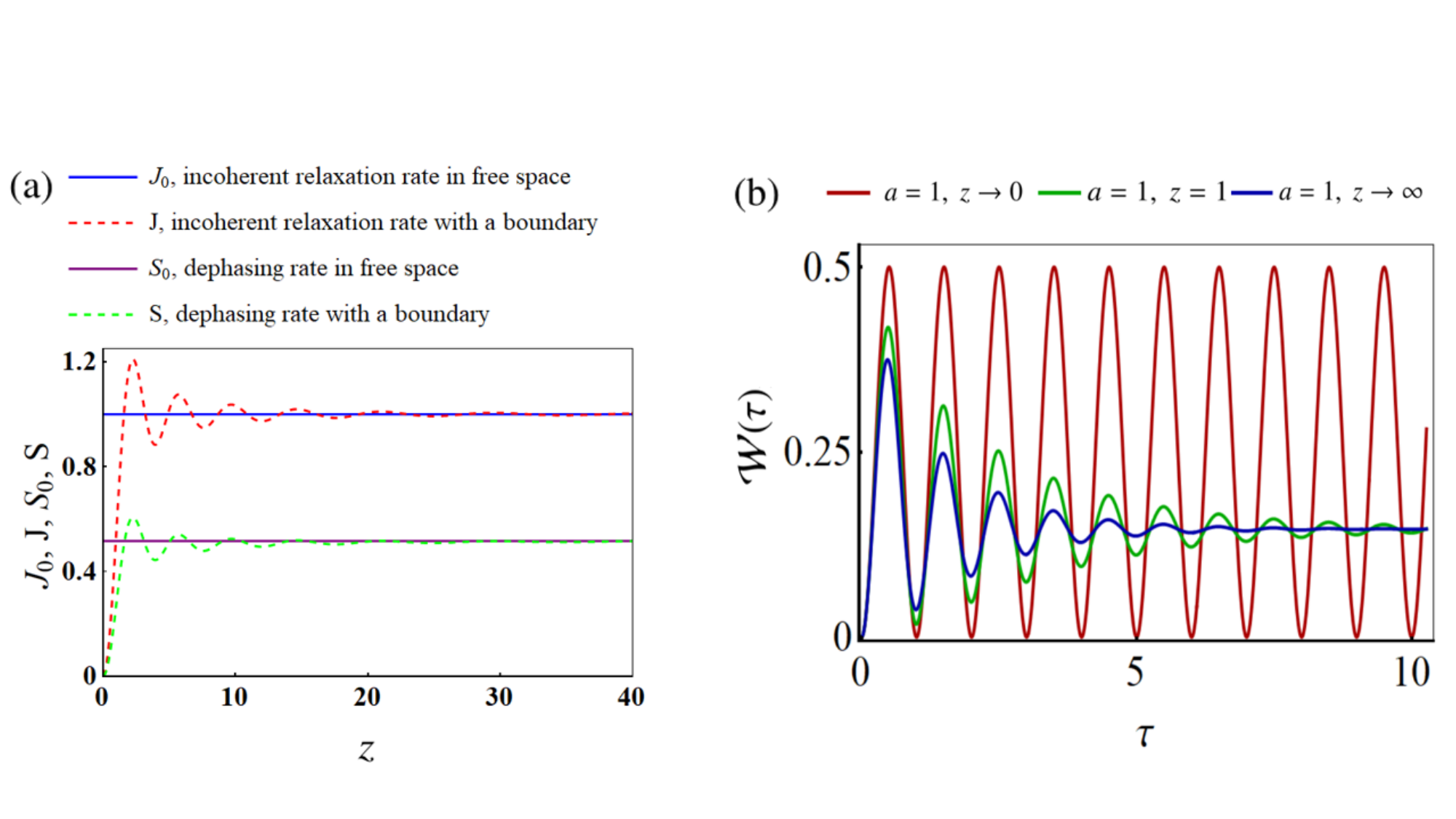}}
\caption{(a) The incoherent relaxation rate and the dephasing rate in free space, i.e., $J_0$ and $S_0$, and the incoherent relaxation rate and the dephasing rate in a space with a boundary, i.e., $J$ and $S$, are plotted as a function of the distance $z$ with $a=0.1$.
(b) Ergotropy $\mathcal{W}(\tau)$ (per unit $\omega_0$) versus time $\tau$ with $a=1$ in a space with a reflecting plane boundary. Here, we fixed $\omega=1$ in the unit $\omega_0$.
}\label{rates}
\end{figure}

The battery's trajectory can be depicted by
\begin{equation}\label{trajectory}
t(\tau)=\frac{1}{a}\sinh a\tau,\;\;\;x(\tau)=\frac{1}{a}\cosh a\tau,\;\;\;z(\tau)= z,\;\;\; y(\tau) = 0.
\end{equation}
The introduction of the reflecting boundary alters the quantum vacuum structure, leading to modifications in the field's two-point correlation function. With the method of images, the
Wightman function of the vacuum massless scalar field in the
presence of the boundary can be expressed as~\cite{Birrell1984}
\begin{equation}\label{correlated function}
G^+(x(\tau),x(\tau'))=G_0^+(x(\tau),x(\tau'))+G_b^+(x(\tau),x(\tau')),
\end{equation}
where $G_0^+(x(\tau),x(\tau'))$ is the Wightman function in free space already obtained in Eq.~({\ref{correlated function1}}), and $G_b^+(x(\tau),x(\tau'))$ is the correction due to the boundary, given by
\begin{align}
&G_b^+(x(\tau),x(\tau'))\nonumber\\
&=\frac{1}{4\pi^2}
\frac{1}{(t-t'-i\epsilon)^2-(x-x')^2-(y-y')^2-(z+z')^2}.\nonumber
\end{align}
By substituting the QB's trajectory given in eq.~(\ref{trajectory}), the total Wightman function in the presence of the boundary is obtained as
\begin{align}\label{correlated function boundary1}
&G^+(x(\tau),x(\tau'))\nonumber\\
&=-\frac{1}{16\pi^2}\bigg[\frac{a^2}{\sinh^2(\frac{a\Delta\tau}{2}-i\epsilon)}-
\frac{a^2}{\sinh^2(\frac{a\Delta\tau}{2}-i\epsilon)-a^2z^2}
\bigg].\nonumber\\
\end{align}
Performing the Fourier transform and applying contour integration, the modified transition rates for the QB-field interaction take the form of
\begin{align}\label{Fourier transforms}
&{\cal G}(\lambda)=\frac{\mu^2\omega_0^2}{2\pi\Omega^2}\frac{\lambda}{1-e^{-2\pi\lambda/a}}\bigg[1-
\frac{\sin[\frac{2 \lambda}{a}\sinh^{-1}(az)]}{2\lambda z\sqrt{1+a^2z^2} }
\bigg],\\
&{\cal G}(0)=\frac{\mu^2\omega^2 a}{4\pi\Omega^2}\bigg[1-
\frac{\frac{2}{a}\sinh^{-1}(az)}{2 z\sqrt{1+a^2z^2} }
\bigg].
\nonumber
\end{align}
Similarly, we will work with
dimensionless quantities by rescaling position $\tilde{z}=\Omega z$. In the following,
we use the term $z=\tilde{z}$ for convenience. Then we can obtain the modified coefficients of the Kossakowski matrix $a_{ij}$:
\begin{align}\label{ABa2}
&A=\frac{\Gamma\omega_0^2}{2\Omega^2}\coth\bigg(\frac{\pi}{a}\bigg)\bigg[1-\frac{\sin f(a,z)}{2z\sqrt{N(a,z)}}\bigg],\nonumber\\
&B=\frac{\Gamma\omega_0^2}{2\Omega^2}\bigg[1-\frac{\sin f(a,z)}{2z\sqrt{N(a,z)}}\bigg],\\
&C=\frac{\Gamma \omega_0^2}{2\Omega^2}\bigg\{\frac{ \omega^2a}{\omega_0^2\pi}\bigg[1-\frac{ f(a,z)}{2z\sqrt{N(a,z)}}\bigg]\nonumber\\
&\;\;\;\;\;\;\;\;\;\;\;\;\;\;\;\;\;-\coth\bigg(\frac{\pi}{a}\bigg)\bigg[1-\frac{\sin f(a,z)}{2z\sqrt{N(a,z)}}\bigg]\bigg\},\nonumber
\end{align}
where
\begin{align}\label{functions}
&f(a,z)=\frac{2}{a}\sinh^{-1}(az),\\
&N(a,z)=1+a^2z^2.\nonumber
\end{align}
Substituting these expressions into eq.~(\ref{scaledergotrogy0}),
the ergotropy in the presence of a boundary becomes
\begin{align}\label{ergotropyb-0}
\mathcal{W}(\tau)&=\frac{\omega_0}{2}\nonumber\\
&\times\bigg[\sqrt{[-\frac{\omega_0}{\Omega} e^{-J\tau}+\coth^{-1}\bigg(\frac{\pi}{a}\bigg)(e^{-J\tau}-1)]^2
+\frac{\omega^2}{\Omega^2} e^{-2S\tau}}\nonumber\\
&-\frac{\omega_0^2}{\Omega^2} e^{-J\tau}+\frac{\omega_0}{\Omega}\coth^{-1}\bigg(\frac{\pi}{a}\bigg)(e^{-J\tau}-1)\nonumber\\
&-\frac{\omega^2}{\Omega^2} e^{-S\tau}\cos(\Omega\tau)\bigg],
\end{align}
where the modified incoherent relaxation and dephasing rates, respectively, are
\begin{align}\label{rate1}
&J=\frac{2\omega_0^2}{\Omega^2}\coth\bigg(\frac{\pi}{a}\bigg)\bigg[1-\frac{\sin f(a,z)}{2z\sqrt{N(a,z)}}\bigg],\\
&S=\frac{J}{2}+\frac{\omega^2a}{\Omega^2\pi}\bigg[1-\frac{ f(a,z)}{2z\sqrt{N(a,z)}}\bigg].\nonumber
\end{align}
Compared with the free-space case in Eq.~(\ref{rate0}), we find that the modified relaxation and dephasing rates depend not only on the thermal signatures, characterized by $\coth(\pi/a)$ and the additional term $\omega^2a/(\Omega^2\pi)$, but also on the boundary-induced corrections through the normalization factor $N(a,z)$ and the function $f(a,z)$. This additional dependence highlights the role of the boundary in modulating the dissipation dynamics of the QB.
To examine the influence of the boundary on the energy dissipation rates, we compare the incoherent relaxation and dephasing rates in free space ($J_0$ and $S_0$) with those in the presence of a boundary ($J$ and $S$) as functions of $z$ in Fig.~\ref{rates}(a).
In particular, when the QB is placed very far from the boundary ($z\rightarrow\infty$), the curves for $J$ and $S$ coincide with their free-space counterparts ($J_0$ and $S_0$). This behavior arises because, as $z\rightarrow\infty$, the factor $[2z\sqrt{N(a,z)}]^{-1}\rightarrow 0$, leading to the modified relaxation and dephasing rates reducing to
\begin{equation}\label{f0}
J_{z\rightarrow \infty}\rightarrow \frac{2\omega_0^2}{\Omega^2}\coth\bigg(\frac{\pi}{a}\bigg)
,\;\;\;\;\;\;S_{z\rightarrow \infty}\rightarrow\frac{J_{z\rightarrow \infty}}{2}+\frac{\omega^2a}{\Omega^2\pi},
\end{equation}
which recovers to the free space case given by eq.~(\ref{rate0}). This indicates that when the QB is very far from the boundary, the suppression of dissipation becomes negligible, and the ergotropy expression in eq.~(\ref{ergotropyb-0}) reduces to that of the Unruh-DeWitt QB in free space [eq.~(\ref{ergotropy1})]. Additionally, as illustrated in Fig.~\ref{rates}(b), when the QB is positioned far from the boundary, the time evolution of the ergotropy exhibits a damped oscillatory behavior, gradually approaching a steady nonzero value. This pattern arises because the boundary effect in this case becomes negligible and the QB interacts with the external environment as it would in free space, resulting in a gradual reduction in energy extraction over time.
In contrast, when the QB is extremely close to the boundary ($z\rightarrow 0$), one can find $J_{z\rightarrow 0}=2S_{z\rightarrow 0}\rightarrow0$.
It tells us that, as $z\rightarrow 0$, the energy dissipation in the QB can be effectively eliminated. This suggests that the Unruh-DeWitt QB can behave as if it were a closed system, with its charging process almost unaffected by environment-induced decoherence and the Unruh effect.
In this case, the ergotropy is given by
\begin{equation}\label{ergotropy2}
\mathcal{W}(\tau)=\frac{\omega_0}{2}\frac{\omega^2}{\Omega^2}[1-\cos(\Omega\tau)],
\end{equation}
which recovers the result for a closed QB in Eq.~(\ref{ergotrogy-closed}), exhibiting a periodic oscillation with time [can be seen from Fig.~\ref{rates}(b)].
Moreover, similar to Fig.~\ref{figure0}, for any given nonzero distance $z$ from the boundary, the evolution of ergotropy over time presents a damped oscillation, whose
amplitude is modulated by the exponential decay dictated by the incoherent relaxation $J$ and dephasing rates $S$. However, when the QB is placed very close to the boundary, it can significantly suppress both incoherent relaxation and dephasing rates, leading to enhanced energy extraction. This suppression effect is attributed to the interference between the original field modes and their reflections from the boundary, effectively modifying the local density of vacuum fluctuations and altering the QB's decoherence dynamics.

Given the dominant influence of the energy dissipation rates on the ergotropy dynamics of the QB, it is essential to provide further insight into the study of their behavior near a boundary under the two limiting conditions, that is, $a z\ll 1$ and $az\gg 1$. Therefore, we identify a characteristic length scale, $z_a=1/a$.
When the distance between the QB and the boundary is very small, that is, the short distance regime $z\ll z_a$, the rates are given by
\begin{equation}\label{rates-smaller}
J_{z\ll z_a}= 2S_{z\ll z_a}\simeq \frac{2\omega_0^2}{\Omega^2}\coth\bigg(\frac{\pi}{a}\bigg)\bigg[1-\frac{\sin(2  z)}{2  z}\bigg],
\end{equation}
which clearly shows that the incoherent relaxation rate and dephasing rate
are characterized by a $z^{-1}$ power-law distance dependence, and increase in an oscillatory manner to that for the free space case [can be seen from Fig.~\ref{rates}(a)], as well as the only net effect of acceleration is embodied in the Unruh thermal analogy.
However, when the distance is very large, that is, $z\gg z_a$, one has
\begin{align}\label{rates-larger}
&J_{z\gg z_a}\simeq \frac{2\omega_0^2}{\Omega^2}\coth\bigg(\frac{\pi}{a}\bigg)\bigg\{1- \frac{1}{2a z^2}\sin\bigg[\frac{2}{a} \ln (2az)\bigg]\bigg\},\\
&S_{z\gg z_a}\simeq \frac{J_{z\gg z_c}}{2}+\frac{a\omega^2}{\Omega^2\pi} \bigg[1-\frac{1}{z^2a^2} \ln (2az)\bigg].\nonumber
\end{align}
Here, the rates exhibit a $z^{-2}$ power-law distance dependence resulting from the signatures of the relativistic acceleration. As illustrated in Fig.~\ref{rates}(a), the oscillations diminish significantly, and the incoherent relaxation and dephasing rates asymptotically approach their corresponding values in free space.

\begin{figure}[H]
\centering
{\includegraphics[height=2.05in,width=3.35in]{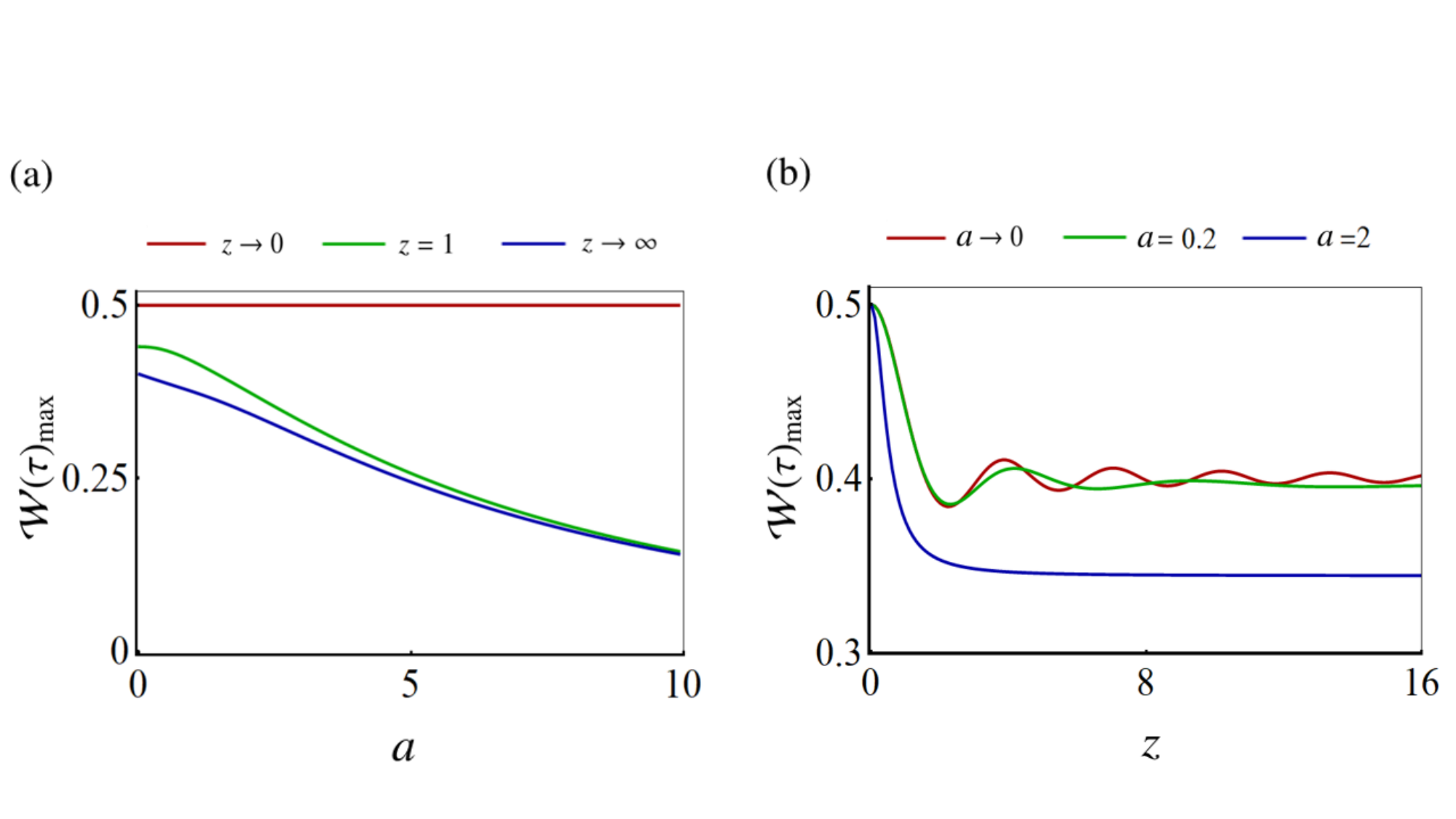}}
\caption{The maximum ergotropy $\mathcal{W}(\tau)_{\mathrm{max}}$  (per unit $\omega_0$) during the evolution time versus (a) acceleration $a$ and (b) distance between the QB and the boundary $z$ in a space with a boundary. Here, we fixed $\omega=1$ in the unit $\omega_0$.
}\label{reflecting1}
\end{figure}

From the above analysis, the boundary plays a crucial role in modulating the QB performance by modifying the vacuum fluctuations, which in turn influence the energy dissipation rates.
By tuning the acceleration and distance parameters ($a$ and $z$), one can effectively control the amount of energy that can be extracted by the QB.
Figure~\ref{reflecting1}(a) illustrates that the maximum ergotropy, $\mathcal{W}(\tau)_{\mathrm{max}}$, decreases with increasing $a$, but this decay is less pronounced when the QB is closer to the boundary. In the extreme limit $z\rightarrow 0$, the ergotropy is effectively shielded from both environment-induced decoherence and accelerated motion, resulting in negligible dissipation [also can be seen from Fig.~\ref{reflecting1}(b)].
Moreover, Fig.~\ref{reflecting1}(b) plots $\mathcal{W}(\tau)_{\mathrm{max}}$ as a function of $z$ for different accelerations.
In the limit $a\rightarrow0$, the energy dissipation rates are approximately given by $J_{a\rightarrow 0}=2S_{a\rightarrow 0}\approx \frac{2\omega_0^2}{\Omega^2}[1- \sin(2z)/(2z)]$, leading to an oscillatory decrease in the maximum ergotropy with increasing $z$.
As the acceleration increases, the oscillations are gradually suppressed. For sufficiently large accelerations, the ergotropy rapidly converges to a steady value.

\subsection{Ergotropy of static QB in a thermal bath with a reflecting boundary}\label{boundaryspace2}
Through the previous discussion, we find that the maximal amount of energy can be extracted from the Unruh-DeWitt QB coupled to a massless scalar field in a space with a reflecting boundary. There is a natural issue of concern, namely the equivalence of the effects of uniform acceleration and the effects of a thermal bath on the energy dissipation.
It is also shown in Refs.~\cite{Unruh1976,Fulling1973,Davies1975,Birrell1984} that in free space, there exists an equality between the behavior of a uniformly accelerated two-level detector and the behavior of an inertial detector in a thermal bath, as we studied in Sec.~\ref{emptyspace}. Then, a natural question is: Is this equivalence still valid for the QB in a space with a reflecting boundary? In the following,
we address this issue in terms of the ergotropy of a QB.

We set the static QB placed at $z$ from the boundary so that the trajectories are given by
\begin{equation}\label{static trajectories}
t(\tau) =\tau,\;\;\;\; x(\tau) = 0,\;\;\;\; z(\tau) = z,\;\;\;\;
y(\tau) = 0.
\end{equation}
It is worth stressing that the Wightman function in a space with a reflecting boundary for a thermal massless scalar field reads~\cite{Birrell1984}
\begin{align}\label{M correlated function}
&G^+(x(\tau),x(\tau'))=-\frac{1}{4\pi^2}\nonumber\\
&\;\;\;\times\sum_{n=-\infty}^{+\infty}\bigg[
\frac{1}{(\Delta\tau-in/T-i\epsilon)^2}+\frac{1}{(\Delta\tau-in/T-i\epsilon)^2-(2z)^2}
\bigg].\nonumber\\
\end{align}
Inserting the Wightman function (\ref{M correlated function}) into the Fourier transforms (\ref{transforms0}), and performing the integrations, we arrive at the following coefficients of $a_{ij}$:
\begin{align}\label{ABT2}
&A=\frac{\Gamma\omega_0^2}{2\Omega^2}\coth\bigg(\frac{1}{2T}\bigg)\bigg[1-\frac{\sin(2  z)}{2  z}\bigg],\nonumber\\
&B=\frac{\Gamma\omega_0^2}{2\Omega^2}\bigg[1-\frac{\sin(2  z)}{2  z}\bigg],\;\;\;\;C=-A.
\end{align}
As a result, the incoherent relaxation and dephasing rates are given by
\begin{equation}\label{ratesT}
J_{T}= 2S_{T}= \frac{2\omega_0^2}{\Omega^2}\coth\bigg(\frac{1}{2T}\bigg)\bigg[1-\frac{\sin(2  z)}{2  z}\bigg].
\end{equation}
The application of these decay rates to Eq.~(\ref{scaledergotrogy0}) gives rise to the ergotropy of the static QB in a thermal bath with a reflecting boundary, which is
\begin{align}\label{ergotropyT}
\mathcal{W}(\tau)
&=\frac{\omega_0}{2}\times\nonumber\\
&\bigg[\sqrt{[-\frac{\omega_0}{\Omega} e^{-J_{T}\tau}+\coth^{-1}\bigg(\frac{1}{2T}\bigg)(e^{-J_{T}\tau}-1)]^2
+\frac{\omega^2}{\Omega^2} e^{-J_{T}\tau}}\nonumber\\
&-\frac{\omega_0^2}{\Omega^2} e^{-J_{T}\tau}+\frac{\omega_0}{\Omega}\coth^{-1}\bigg(\frac{1}{2T}\bigg)(e^{-J_{T}\tau}-1)\nonumber\\
&-\frac{\omega^2}{\Omega^2} e^{-\frac{J_{T}}{2}\tau}\cos(\Omega\tau)\bigg].
\end{align}

\begin{figure}[H]
\centering
{\includegraphics[height=1.95in,width=3.35in]{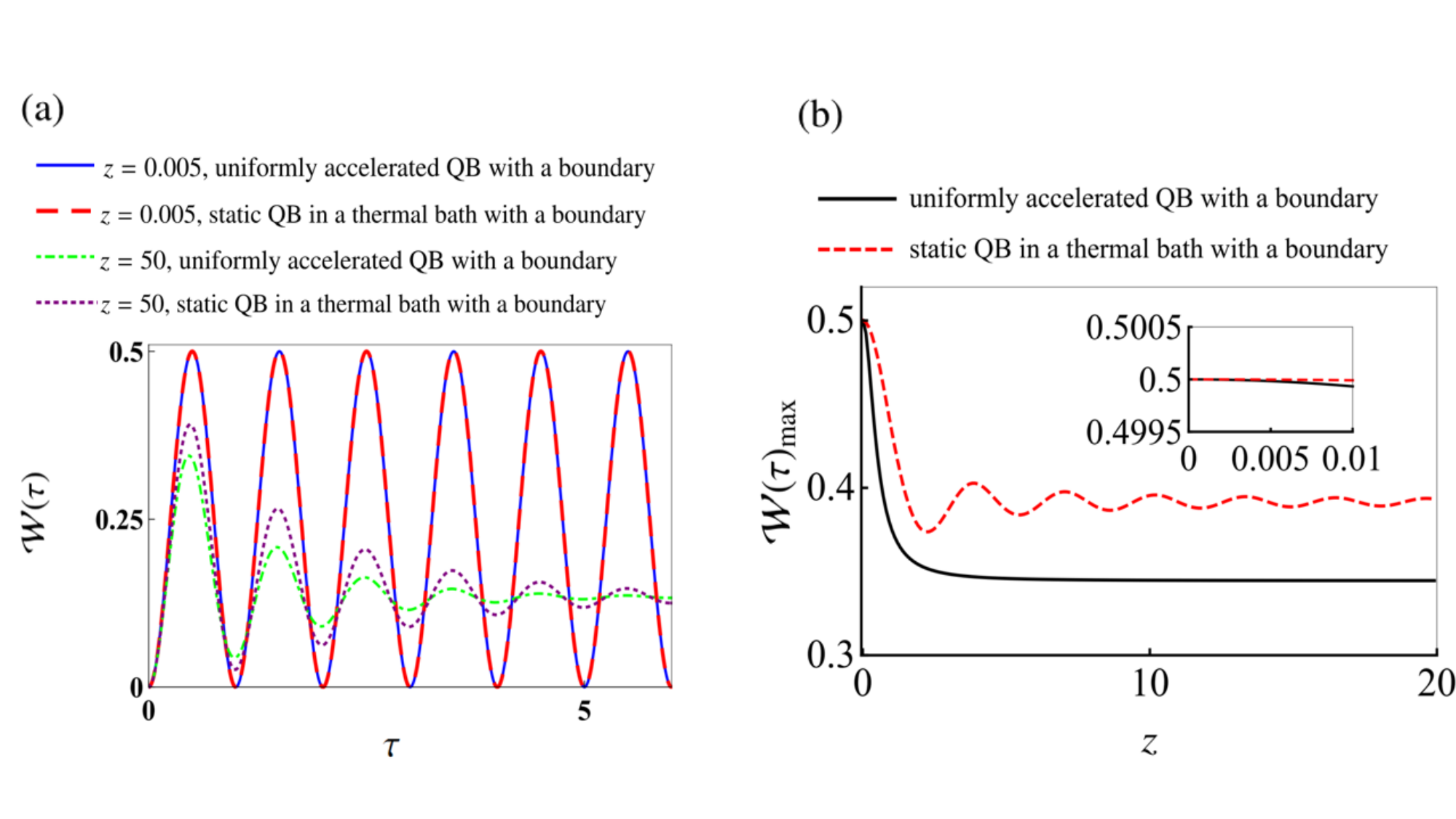}}
\caption{(a) Ergotropy $\mathcal{W}(\tau)$ (per unit $\omega_0$) versus time $\tau$ with different values $z=0.005, 50$, and (b) the maximum ergotropy $\mathcal{W}(\tau)_{\mathrm{max}}$  (per unit $\omega_0$) with time versus the distance $z$, for a uniformly accelerated QB and for a static QB in a thermal bath, respectively. Here, we fixed $z_a=0.5$ and $\omega=1$ in the unit $\omega_0$.
}\label{compare-a}
\end{figure}

To analyze the boundary effects on the ergotropy of a static QB in a thermal bath at temperature $T=a/2\pi$, and to compare these results with those of an accelerated QB, we plot the time evolution of the ergotropy $\mathcal{W}(\tau)$ (per unit $\omega_0$) in Fig.~\ref{compare-a}. Here, we fixed the parameters $z_a=0.5$ and $z=0.005,\;50$, respectively, in Fig.~\ref{compare-a}(a). We find that when the distance from the boundary is much smaller than the characteristic length, that is, $z\ll z_a$, the ergotropy curve for the accelerated QB nearly coincides with that for the static QB in a thermal bath at the corresponding Unruh temperature. This indicates that, in the regime $z\ll z_a$, the dissipative dynamics of the accelerated QB is effectively indistinguishable from those of a static QB in a thermal bath. In contrast, for $z\gg z_a$, the time-dependent behavior of the ergotropy differs between these two scenarios, indicating markedly different dissipative dynamics.

Moreover, to further illustrate the position-dependent suppression of dissipation induced by the presence of a boundary, we examine the maximum ergotropy over time as a function of the distance $z$ in Fig.~\ref{compare-a}(b). Here, we fixed $z_a=0.5$.
This figure shows that as $z$ decreases, the suppression of dissipation becomes stronger.
We can see from the inset that, when $z\ll z_a$, the suppression of dissipation is exceptionally robust. In addition, the maximum ergotropy of accelerated QB approaches that of static QB in a thermal bath. This behavior can be understood from the expressions for the dissipation rates. In the regime $z\ll z_a$, for the accelerated QB, the decay rates are given by Eq.~(\ref{rates-smaller}): $J_{z\ll z_a}= 2S_{z\ll z_a}\simeq \frac{2\omega_0^2}{\Omega^2}\coth(\frac{\pi}{a})[1-\sin(2  z)/(2  z)]$,
and for the static QB in a thermal bath at temperature $T=a/2\pi$, the rates are given by Eq.~(\ref{ratesT}): $J_{T}= 2S_{T}= \frac{2\omega_0^2}{\Omega^2}\coth(\frac{1}{2T})[1-\sin(2  z)/(2  z)]$.
Thus, in the limit $z\ll z_a$, the equivalence $J_{z\ll z_a}=J_{T=a/2\pi}$ is maintained,
implying that the boundary-induced suppression of dissipation for the accelerated QB is essentially similar to that for a static QB in a thermal bath at temperature $T=a/2\pi$.
However, as $z$ increases, the suppression effect weakens and the ergotropy of the accelerated QB declines more rapidly than that of the static QB in a thermal bath, as illustrated in Fig.~\ref{compare-a}(b). This difference arises from the distinct decay patterns of the energy dissipation rates. For the accelerated QB in the regime $z\gg z_a$, the rates are given by Eq.~(\ref{rates-larger}), i.e., $J_{z\gg z_a}\simeq \frac{2\omega_0^2}{\Omega^2}\coth(\frac{\pi}{a})[1- \frac{1}{2a z^2}\sin[\frac{2}{a} \ln (2az)]]$ and
$S_{z\gg z_a}\simeq \frac{J_{z\gg z_c}}{2}+\frac{a\omega^2}{\Omega^2\pi} [1-\frac{1}{z^2a^2} \ln (2az)]$, compared to those for the static QB in a thermal bath in Eq.~(\ref{ratesT}). In particular, the incoherent relaxation and dephasing rates for the accelerated QB vanish more quickly with power law $1/z^2$, while for the static one in a thermal bath case, the incoherent relaxation and dephasing rates fade away in an oscillatory with power law $1/z$.
Therefore, in the limit $z\gg z_a$, the boundary-induced suppression of dissipation for the accelerated QB is significantly different from that of the static QB in a thermal bath at temperature $T=a/2\pi$.

These results tell us that the characteristic length scale $z_a$ marks as a crossover length that distinguishes regimes dominated by boundary with regimes dominated by acceleration. For $z\ll z_a$, the boundary-induced modification of field correlations becomes dominant. The reflected field modes constructively interfere with the incoming vacuum modes, leading to a strong suppression of vacuum fluctuations. It is possible to define a local inertial frame in which the boundary-induced suppression of dissipation for the accelerated QB is well approximated by its thermal Minkowski analogue. For $z\gg z_a$, the boundary-induced effect weakens, and the non-inertial character of relativistic acceleration becomes dominant, encoded in the non-Minkowskian metric. In this limit, the non-inertial geometry and spacetime non-locality amplify the difference between its fluctuations and thermal fluctuations, causing the boundary-induced suppression of dissipation for the accelerated QB to differ significantly from that for a static QB in a thermal bath. This scale $z_a$   corresponds precisely to the breakdown of the local thermal approximation, as established in relativistic quantum systems~\cite{Misner1973,Marino2014}.


\section{Conclusions} \label{sectionIV}
In conclusion, we have examined the dynamics of an Unruh-DeWitt QB interacting with the vacuum fluctuations of a massless scalar field, both in free space and in the presence of a reflecting boundary. We introduced an external classical driving force as the charger and modeled the QB-field coupling in the transverse direction.
Our results demonstrate that the environment-induced decoherence leads to the energy dissipation of the QB, and this effect is significantly enhanced by accelerated motion. In free space, the dissipative behavior of the accelerated QB is equivalent to that of a static QB in a thermal bath at temperature $T=a/2\pi$, consistent with the Unruh effect.

However, when a reflecting boundary is introduced, the dissipation suppression becomes position-dependent. As the QB approaches the boundary, we can effectively suppress the energy dissipation in the QB caused by environment-induced decoherence and the Unruh effect. In particular, when the QB is placed very close to the boundary, energy dissipation is nearly eliminated, as if the QB were an isolated system. Furthermore, there exists a characteristic distance $z_a$. When the distance $z$ between the QB and the boundary is smaller than $z_a$, that is, $z\ll z_a$, the dissipation suppression is highly effective, and the suppression effect for the accelerated QB is indistinguishable from that for a static QB in a thermal bath.
In contrast, when the distance exceeds $z_a$, the ergotropy of the accelerated QB decreases rapidly, while for a static QB in a thermal bath at temperature $T=a/\pi$, it decays following an oscillatory pattern.
This indicates that for $z\gg z_a$, the suppression effect weakens, and the suppression effect for the accelerated QB differs significantly from that for a static QB in a thermal bath.

Recent studies on quantum thermal machines, such as heat engines and refrigerators, focus on cyclic thermodynamic protocols that involve energy exchange with thermal reservoirs~\cite{Binder2018,Myers2022}. Key performance metrics include optimizing work output, efficiency, cooling power, and entropy production~\cite{Mitchison2019,Uzdin2015,Ghosh2018}. In contrast, QBs are typically modeled as quantum systems (e.g., two-level systems, spin chains, Dicke models) that store and release energy through external driving or field interactions, without relying on cyclicity or temperature gradients~\cite{Alicki2013,Campaioli2023,Campaioli2018}. Their performance is characterized by metrics such as charging power, extractable energy,
and resistance to dissipation~\cite{Binder2015,Andolina2019,Hovhannisyan2013,Shi2022,Gyhm2024,Ferraro2018,Campaioli2017,Riccardo2024}.
One of the common quantities characterizing these quantum machines is ergotropy, which quantifies the maximum extractable work from a given quantum state~\cite{Allahverdyan2004,Francica2020,Cakmak2020}. However, its physical origin differs: in thermal machines, ergotropy arises from population inversion or coherence induced by thermal reservoirs, whereas in QBs it reflects stored energy that can be degraded by dissipation.
Our investigations provide valuable insights for optimizing the performance of quantum machines in practical scenarios. Using boundary effects strategically, it is possible to effectively address challenges such as environment-induced decoherence and energy dissipation.
The boundary-induced suppression mechanism relies on the modification of field correlations, which determine the transition rates of the QB. This mechanism is expected to remain relevant in multi-qubit systems, where collective decoherence and entanglement may arise~\cite{Merkli2008,Lin2009,Braun2002}, as well as in higher-dimensional spacetimes, where the structure of the Wightman function and Unruh temperature are dimension-dependent~\cite{Hao2025,Feng2022,Yan2022}. In both cases, although the quantitative behavior may change, the suppression of dissipation via boundary-induced vacuum fluctuations modification is expected to persist.
For massive scalar fields, the Wightman function exhibits a qualitatively different behavior~\cite{Zhou2012,Zhou2021,Huang2022,Pan2024}, and the analogy to thermal fluctuations breaks down. Although boundary effects have not yet been fully explored in this context, they are still expected to suppress dissipation over a spatial range determined by the field's mass.
These points could be a promising direction for future research on the development of high-performance relativistic QBs.

\begin{acknowledgments}
This work was supported by the Key Program of the National Natural Science Foundation of China (Grant Nos. 12035005, and 12065016), the
Young Elite Scientist Sponsorship Program by Guizhou Science and Technology Association (Grant No. GASTYESS202424), the Discipline-Team of
Liupanshui Normal University of China (Grant No. LPSSY2023XKTD11),
the Guizhou Provincial Department of Education Higher Education Science Research Project for Youth Project (Grant Nos. Qian Jiao Ji [2022]
345, and Qian Jiao Ji [2022] 346), the Scientific Research Start-Up Funds
of Hangzhou Normal University (Grant No. 4245C50224204016), and
Hangzhou Leading Youth Innovation and Entrepreneurship Team Project
(Grant No. TD2024005).
\end{acknowledgments}





\end{document}